\documentclass[11pt,a4paper]{article}
\usepackage{amsfonts,amssymb,amsmath}
\usepackage{color}
\usepackage{theorem,cite}
\numberwithin{equation}{section}

\definecolor{brique}{rgb}{.9,.2,0}
\definecolor{blvert}{rgb}{0,.8,.85}
\definecolor{vertcl}{rgb}{0,1,.7}
\newcommand\vertcl[1]{\textcolor{vertcl}{#1}}
\newcommand\blvert[1]{\textcolor{blvert}{#1}}
\newcommand\brique[1]{\textcolor{brique}{#1}}
\def\lapth{
\begin{picture}(164,70)(0,-15)\thicklines
\put(0,0){\vertcl{\rule{20pt}{4pt}}}
\put(19,1){\vertcl{\line(1,3){23}}} 
\put(20,1){\vertcl{\line(1,3){23}}} 
\put(21,1){\vertcl{\line(1,3){23}}}
\put(22,1){\vertcl{\line(1,3){23}}}
\put(45,70){\vertcl{\line(1,-3){23}}} 
\put(44,70){\vertcl{\line(1,-3){23}}} 
\put(43,70){\vertcl{\line(1,-3){23}}}
\put(42,70){\vertcl{\line(1,-3){23}}}
\put(2,24){\vertcl{\rule{120pt}{4pt}}}
\put(65,0){\vertcl{\rule{60pt}{4pt}}}
\put(5,37){\Huge{\brique{\textbf{L}}}} 
\put(62,37){\Huge{\brique{\textbf{PTh}}}}
\put(12,-8){\blvert{\rule{92pt}{3.5pt}}}
\put(24,-15){\blvert{\rule{57pt}{3.5pt}}}
\put(36,-22){\blvert{\rule{30pt}{3.5pt}}}
\end{picture}
\raisebox{35pt}{
\begin{minipage}{320pt}\begin{center}
\textbf{Laboratoire d'Annecy-le-Vieux de Physique Th\'eorique}\\[4ex]
website: \texttt{http://lappweb.in2p3.fr/lapth-2005/}
\end{center}
\end{minipage}}\\
\vspace{10pt}\quad \hrulefill\\
\vspace{10pt}}

\newcommand{\be}{\begin{equation}}
\newcommand{\ee}{\end{equation}}
\newcommand{\hs}[1]{\hspace{#1 mm}}
\newcommand{\cc}[1]{c_{#1}^{\phantom{\dagger}}}
\newcommand{\cd}[1]{c_{#1}^{\dagger}}
\newcommand{\nn}[1]{n^{\phantom{\dagger}}_{#1}}
\newcommand{\su}{^{\uparrow}}
\newcommand{\giu}{^{\downarrow}}

                    \def\cL{{\cal L}}

\def\cS{{\cal S}}                    
\def\cV{{\cal V}}          \def\cW{{\cal W}}

\newcommand{\II}{{\mathbb I}}

\newcommand{\MM}{\mbox{${\mathbb M}$}}

\newcommand{\ZZ}{{\mathbb Z}}

\newcommand{\wt}[1]{\widetilde{#1}}
\newcommand{\mb}[1]{\hs{4}\mbox{#1}\hs{4}}

\newcommand{\half}{\frac{1}{2}}

\def\str{\mathop{\rm str}\nolimits}

\theorembodyfont{\rmfamily}

\begin{document}

\thispagestyle{empty}
\renewcommand{\thefootnote}{\fnsymbol{footnote}}

\hspace{-1cm}\lapth
\rightline{LAPTH-Conf-1222/07}

\vspace{20mm}

\begin{center}

{\LARGE{\sffamily Generalised integrable Hubbard models\footnotemark}}\\[1cm]

\footnotetext{Talk given by G. Feverati at the workshop "RAQIS'07 Recent Advances in Quantum Integrable Systems", 11-14 September 2007, LAPTH, Annecy-le-Vieux, France.}
{\large J. Drummond, \underline{G. Feverati}, L. Frappat and E. Ragoucy\footnotemark\\[.7cm] 
\textit{Laboratoire de Physique Th{\'e}orique LAPTH\\  
CNRS, UMR 5108, associ\'e \`a l'Universit\'e de Savoie\\  
9, Chemin de Bellevue, BP 110, F-74941  Annecy-le-Vieux Cedex, France. }}
\end{center}
\vfill

\footnotetext{drummond@lapp.in2p3.fr, feverati@lapp.in2p3.fr, frappat@lapp.in2p3.fr, 
ragoucy@lapp.in2p3.fr}

\begin{abstract}
We construct the XX and Hubbard-like models based on unitary superalgebras $gl(N|M)$ generalizing Shastry's and Maassarani's approach. 

We introduce the R-matrix of the $gl(N|M)$ XX-type model; the one of the Hubbard-like model is defined by "coupling" two independent XX models. In both cases, we show that the R-matrices satisfy the Yang-Baxter equation. We derive the corresponding local Hamiltonian in the transfer matrix formalism and we determine its symmetries. 
 
A perturbative calculation "\`a la Klein and Seitz" is performed. 
Some explicit examples are worked out.
We give a description of the two-particle scattering. 
\end{abstract}

\vfill


\newpage
\renewcommand{\thefootnote}{\arabic{footnote}}
\setcounter{footnote}{0}
\section{Introduction}
The Hubbard model was introduced in order to study strongly correlated
electrons \cite{Hubbard,Gutzwiller} and, since then, it has been widely studied,
essentially due to its connection with condensed matter physics. 
It has been used to describe the Mott
metal-insulator transition \cite{Mott,Hubbard3}, high $T_c$
superconductivity \cite{Anderson,Affleck}, band magnetism \cite{Lieb}
and chemical properties of aromatic molecules \cite{heilieb}.
The
literature on the Hubbard model being rather large, we do not aim at being
exhaustive and rather refer to the books \cite{Monto,EFGKK} and references
therein. Exact results have been mostly obtained in the case of the
one-dimensional model, which enters the framework of our study. In
particular, the 1D model eigenvalues have been obtained by means of the coordinate 
Bethe Ansatz in the celebrated paper by Lieb and Wu \cite{LiebWu}.

One of the main motivations for the present study of the Hubbard model and
its generalisations is the fact that it has recently appeared in the
context of $N=4$ super Yang-Mills theory. Indeed, it
was noticed in \cite{Rej:2005qt} that the Hubbard model at half-filling,
when treated perturbatively in the coupling, reproduces the long-ranged
integrable spin chain of \cite{Beisert:2004hm} as an effective theory. It
thus provides a localisation of the long-ranged spin chain model and
gives a potential solution to the problem of describing interactions
which are longer than the length of the spin chain. The Hamiltonian of this
chain was conjectured in \cite{Beisert:2004hm} to be an all-order
description of the dilatation operator of $N=4$ super Yang-Mills in the
$su(2)$ subsector. That is, the energies of the spin chain are conjectured
to be the anomalous dimensions of the gauge theory operators in this
subsector. 
In relation to this, an interesting approach to the Hubbard model is
given in \cite{ffgr} that leads to the evaluation of energies for the
antiferromagnetic state and allows one to control the order of the limits
of large coupling and large length of the operators/large angular momentum.

There may be the possibility that some integrable extension of the Hubbard
model could be put in relation to other subsectors of 
the $N=4$ super Yang-Mills theory. Here we
will discuss a general approach to constructing a number of supersymmetric
Hubbard models. Each of these models can be treated perturbatively and thus
gives rise to an integrable long-ranged spin chain as an effective theory.

Other supersymmetric generalisations of the Hubbard model have been
constructed, see e.g. \cite{EKS,BGLZ}. These approaches mainly concern high
$T_c$ superconductivity models and their relation with the $t-J$ model.
They essentially use the $gl(1|2)$ or $gl(2|2)$ superalgebras, which appear
as the symmetry algebras of the Hamiltonian of the model. Our approach
however is different and is based on the QISM framework. It ensures the
integrability of the model and allows one to obtain local Hubbard-like
Hamiltonians for general $gl(N|M)$ superalgebras. They can be interpreted in
terms of `electrons' after a Jordan--Wigner transformation.

In this review paper we revisit and slightly extend the results of \cite{DFFR1}, our goal
here being not to reproduce the calculations but to focus on the main
ideas of our approach.

The plan of the paper is as follows. 
Section~\ref{sect:Hubbard} is devoted to sketch a number of facts for the ordinary Hubbard model.
In section \ref{sect:XX}, we define universal XX models. 
We introduce the corresponding Hamiltonians and determine the
symmetries of the model. 
In section \ref{sect:Hub}, we summarise  the construction of the
associated Hubbard-like model, in the Shastry and Maassarani approach. 
From the transfer matrix we obtain the Hamiltonian; we also discuss the symmetries.
In section~\ref{sect:pert} we perform a second order perturbative computation \textit{\`a la} Klein and Seitz \cite{klstz}. 
Then, we define the Jordan-Wigner transformation, section~\ref{sect:JW}, used in 
section \ref{sect:examples} to give some examples where we write explicitly the
Hamiltonians in the $gl(2|2)$, $gl(4)$ and $gl(4|4)$ cases. 
We finish in section~\ref{twopart} with a study of two-particle interactions.

\section{Hubbard model\label{sect:Hubbard}}
The 1 dimensional Hubbard model introduced by \cite{Hubbard,Gutzwiller}
describes hopping electrons on a lattice, with an ultralocal repulsive potential 
that implements a screened Coulomb repulsion, with $U>0$. The Hamiltonian is given by
\be\label{oldHubb}
H=-t \sum _{i=1}^L \sum _{\rho=\uparrow , \downarrow}
\left(e^{i\phi} \cd{\rho,i}\cc{\rho,i+1}+e^{-i\phi}\cd{\rho,i+1}\cc{\rho,i}\right) 
+U \sum_{i=1}^L \big(1-2\nn{\uparrow,i}\big)\big(1-2\nn{\downarrow,i}\big)
\ee
We will always use periodic boundary conditions. 

In $\mathcal{N}=4$-SYM theory this model was first observed in \cite{Rej:2005qt}, where
a magnetic flux $\phi$ of Aharonov-Bohm type was included. In that paper, 
the potential term was written in a slightly different but equivalent form.
The relation between couplings was identified and the system was taken at half-filling;
for our needs we just observe that the ratio $t/U$ corresponds to the coupling $g$
\be\label{couplings}
\frac{t}{U}=\frac{g}{\sqrt{2}}\,.
\ee
We observe that the Hamiltonian is Hermitian if $\phi\in\mathbb{R}$. 
In the following we will work with this flux equal to zero.

The underling algebraic structure leads us to superalgebras: on each site $i$ the 
fermionic structure 
\be
\{\cc{\rho,i},\cd{\rho',j}\}=\delta_{\rho,\rho'}\delta_{i,j} \, \qquad
\{\cc{\rho,i},\cc{\rho',j}\}=\{\cd{\rho,i},\cd{\rho',j}\}=0
\ee
is a realisation of the super-Lie algebra $gl(1|1)\oplus gl(1|1)$.
The full model algebra is obtained by $L$-times the tensor product of the one site structure. 
We can easily represent the fermionic structure by a graded tensor product of Pauli matrices,
written here with the standard notation for basis matrices $E_{\alpha\beta}$ to emphasise
the grading:
\begin{gather}
E_{12;\rho,i}=\cc{\rho,i} \,,\quad  
E_{21;\rho,i}=\cd{\rho,i} \,,\quad
E_{22;\rho,i}=\nn{\rho,i}=\cd{\rho,i}\cc{\rho,i} \,,\quad
E_{11;\rho,i}=1-\nn{\rho,i}=\cc{\rho,i}\cd{\rho,i} \label{JW}\\[3mm]
E_{12}=\begin{pmatrix} 0 & 1\\ 0 & 0\end{pmatrix} \,,\qquad 
E_{21}=\begin{pmatrix} 0 & 0\\ 1 & 0\end{pmatrix} \,,\qquad
E_{11}=\begin{pmatrix} 1 & 0\\ 0 & 0\end{pmatrix}\,,\qquad
E_{22}=\begin{pmatrix} 0 & 0\\ 0 & 1\end{pmatrix}\,. \notag
\end{gather}
When it occurs, the second pair of labels $\rho,i$ indicates the spin polarisation $\rho$ 
and the site $i$.
The matrices $E_{12}\,,E_{21}$ are taken of fermionic character (they satisfy anticommutation relations whatever their spin and space labels are) and $E_{11}\,,E_{22}$ are taken of bosonic character (they always enter commutation relations whatever their spin and space labels are).
The relation (\ref{JW}) is a graded Jordan-Wigner transformation\footnote{The ordinary
Jordan-Wigner transformation is 
$\displaystyle\cd{\uparrow,i}=\sigma^{-}_{\uparrow,i} \prod _{k>i} \sigma^{z}_{\uparrow,i}$
for the up polarisation; an additional term occurs for the down polarisation.} 
and respects periodic boundary conditions\footnote{The standard one violates periodicity.}.
We now rewrite the Hamiltonian in the spin chain language
\begin{eqnarray} 
H &= &-t \sum _{i=1}^L \sum _{\rho=\uparrow , \downarrow}
\left( E_{21;\rho,i}\ E_{12;\rho,i+1}+E_{21;\rho,i+1}\ E_{12;\rho,i}\right) 
+U \sum_{i=1}^L  
\big(E_{11;\uparrow,i}-E_{22;\uparrow,i}\big)\big(E_{11;\downarrow,i}-E_{22;\downarrow,i}\big) \nonumber 
\end{eqnarray}
and we split it into the sum of the two polarisations 
\begin{gather}
H= H_{\text{XX}}\su +H_{\text{XX}}\giu +U \sum_{i=1}^L 
\big(E_{11;\uparrow,i}-E_{22;\uparrow,i}\big)\big(E_{11;\downarrow,i}-E_{22;\downarrow,i}\big)\;; \label{hubbspin}\\
H_{\text{XX}}^{\rho} = -t \sum _{i=1}^L 
\left( E_{21;\rho,i}\ E_{12;\rho,i+1}+E_{21;\rho,i+1}\ E_{12;\rho,i}\right)\;. \nonumber
\end{gather}
Taking one polarisation of the kinetic term we easily see that
\begin{gather}
E_{21;\rho,i}\ E_{12;\rho,i+1}+E_{21;\rho,i+1}\ E_{12;\rho,i}=
\frac12 \Big[ E_{x;\rho,i} \ E_{x;\rho,i+1}+ E_{y;\rho,i}\  E_{y;\rho,i+1} \Big]\\[3mm]
E_{x;\rho,i}=\begin{pmatrix} 0 & 1\\ 1 & 0\end{pmatrix}_{\rho,i} \,,\quad 
E_{y;\rho,i}=\begin{pmatrix} 0 & -i\\ i & 0\end{pmatrix}_{\rho,i} \notag
\end{gather}
namely we see the appearance of a (graded) XX spin chain Hamiltonian\footnotemark\
(or better two XX spin chains, one for each polarisation) within the Hubbard model. 
\footnotetext{At this point it should be clear that the difference between graded and non
graded cases appears when boundary effects are observed; the thermodynamic limit usually ignores such terms, being sensitive to bulk contributions only.} 

It turns out that the breaking of (\ref{hubbspin}) into the Hamiltonian of 
two XX models plus a potential will allow us to generalise this model to higher 
algebraic structures by maintaining its main property: 
integrability\footnote{The flux $\phi$ does not affect integrability properties.}.

A first hint of integrability of the Hubbard model 
came from the coordinate Bethe Ansatz solution obtained by 
Lieb and Wu \cite{LiebWu} but a full understanding of it by the existence of an infinite 
set of commuting charges came much later. A complete set of eigenstates was constructed in
\cite{EKScomp} using the $SO(4)$ symmetry of the 1D Hubbard Hamiltonian.
Within the framework of
the quantum inverse scattering method, an R-matrix was first constructed by Shastry
\cite{shastry,JWshas} and Olmedilla et al. \cite{Akutsu}, by coupling
(decorated) R-matrices of two independent $XX$ models, through a term
depending on the coupling constant $U$ of the Hubbard potential. 
The proof of the Yang--Baxter relation for the R-matrix was given by
Shiroishi and Wadati \cite{shiro2}. 
With a standard construction, a transfer matrix can be constructed by taking the trace of 
a tensor product of R-matrices. The Yang-Baxter equation guarantees that the transfer matrix 
is the generating functional of an infinite set of commuting charges. 
One of these charges is the Hamiltonian (\ref{hubbspin}) itself.

The construction of the
R-matrix was then generalised in the $gl(N)$ case by Maassarani et al.,
first for the XX model \cite{maasa} and then for the $gl(N)$ Hubbard model
\cite{maasa2,maasa3}. Within the QISM framework, the eigenvalues of the
transfer matrix of the Hubbard model were found using the algebraic
Bethe Ansatz together with certain analytic properties in
\cite{YueDegu,RamMar,martins}.

\section{Universal XX models\label{sect:XX}}
We generalize the construction given in \cite{maasa,martins,DFFR1} to the
case of an arbitrary representation space $\cV$, possibly infinite
dimensional. 
We will use the standard auxiliary space notation, i.e. to any operator 
$A\in \mbox{End}(\cV)$, we associate the operator $A_{1}=A\otimes \II$ and
$A_{2}=\II\otimes A$ in $\mbox{End}(\cV)\otimes \mbox{End}(\cV)$. More
generally, when considering expressions in $\mbox{End}(\cV)^{\otimes k}$,
$A_{j}$, $j=1,\ldots,k$ will act trivially in all spaces $\mbox{End}(\cV)$, but
the $j^{th}$ one.

To deal with superalgebras, we will also need a $\ZZ_{2}$ grading $[.]$ on
$\cV$, such that $[v]=0$ will be associated to bosonic states
$v\in\cV$ and $[v]=1$ to fermionic ones. 

We will also assume the existence of a (super-)trace operator, defined on a subset of 
$\mbox{End}(\cV)$ and obeying cyclicity. When $\cV$ is finite dimensional, 
$\mbox{dim}(\cV)=K$, $\mbox{End}(\cV)$ is a matrix algebra or super-algebra so that 
the trace operator is the usual trace or supertrace of $K\times K$ matrices. 
When $\cV$ is infinite dimensional, the definition of a trace 
operator is more delicate and we will just assume that it exists and is cyclic, 
for the operators we use. 

The construction of a universal XX model is mainly based on general properties 
of a given projector and a permutation. Our main projectors are chosen in $\mbox{End}(\cV)$
as being 
\begin{eqnarray}
\pi:\ \cV\to\ \cW\quad,\quad 
\wt\pi=\II-\pi:\  \cV\to\ \wt{\cW} \mbox{~~~with~~~} \cV=\cW\oplus\wt{\cW}
\label{def:univpi}
\end{eqnarray}
In the tensor product of two vector spaces we take the (possibly graded) permutation
\begin{eqnarray}
P_{12}:
\begin{cases} \cV\otimes\cV \ \to\ \cV\otimes\cV\\
v_{1}\otimes v_{2}\ \to\ (-1)^{[v_{1}][v_{2}]}\, v_{2}\otimes v_{1}
\end{cases}
\end{eqnarray}
For example, in the superalgebra $gl(N|M)$ a possible choice is
\be \label{example}
\displaystyle \pi=\sum_{j \neq N,N+M} E_{jj}\quad,
\quad \wt{\pi}=\II-\pi = E_{NN}+E_{N+M,N+M}
\ee

\subsection{R-matrix\label{sec:univR-XX}}
From the previous operators, one can construct an R-matrix acting on $\cV\otimes\cV$
\begin{equation}
R_{12}(\lambda) = \Sigma_{12}\,P_{12} + \Sigma_{12}\,\sin\lambda +
(\II\otimes\II-\Sigma_{12})\,P_{12}\,\cos\lambda
\label{def:univRXX}
\end{equation}
where $\Sigma_{12}$ is built on the projection operators:
\begin{eqnarray}
\Sigma_{12} &=& 
\pi_{1}\,\wt\pi_{2}+\wt\pi_{1}\,\pi_{2} 
\label{def:univSigma}
\end{eqnarray}
It is easy to show that $\Sigma_{12}$ is also a projector in
$\cV\otimes\cV$:
$\left(\Sigma_{12}\right)^2=\Sigma_{12}$.

Let us introduce the operator $C$:
\begin{equation}
C = \pi-\wt\pi\,.
\label{eq:opC}
\end{equation}
It obeys $C^{2}=\II$ and is related to the R-matrix through the 
equalities
\begin{equation}
\Sigma_{12}=\half(1-C_{1}C_{2}) \mb{and}
\II\otimes\II-\Sigma_{12}=\half(1+C_{1}C_{2})
\label{eq:univSig-C}
\end{equation}
In \cite{DFFR1} we gave proof of a number of useful properties of the R-matrix. 
Essentially the same proofs work also for the slightly more general 
formulation given here. 
The main properties are unitarity, regularity, and Yang--Baxter equation (YBE), that guarantees us that we have an integrable model
\begin{gather}
R_{12}(\lambda_{12})\,R_{13}(\lambda_{13})\,R_{23}(\lambda_{23}) = 
R_{23}(\lambda_{23})\,R_{13}(\lambda_{13})\,R_{12}(\lambda_{12})
\qquad\notag \\[2mm]
\mb{where} \lambda_{ij} = \lambda_i-\lambda_j.
\label{eq:univYBE}
\end{gather}

\subsection{Monodromy and transfer matrix\label{transferXX}}
With a very standard construction, from the R-matrix one constructs the ($L$ sites) 
monodromy matrix
\begin{equation}
\cL_{0<1\ldots L>}(\lambda) = R_{01}(\lambda)\,R_{02}(\lambda)\cdots
R_{0L}(\lambda)
\end{equation}
where we tensor product one R-matrix for each site of the theory. 
It obeys the relation
\begin{equation}
R_{00'}(\lambda-\mu)\, \cL_{0<1\ldots L>}(\lambda)\, 
\cL_{0'<1\ldots L>}(\mu) =
\cL_{0'<1\ldots L>}(\mu) \, \cL_{0<1\ldots L>}(\lambda)\,
R_{00'}(\lambda-\mu)\,.
\label{univRLL-XX}
\end{equation}
where $0$ and $0'$ are two copies of the auxiliary space. 
This relation allows us to construct an ($L$ sites) integrable XX spin
chain through the transfer matrix
\begin{equation}
t_{1\ldots L}(\lambda) = \str_{0} \cL_{0<1\ldots L>}(\lambda) 
= \str_{0} \Big(R_{01}(\lambda)\,R_{02}(\lambda)\cdots 
R_{0L}(\lambda)\Big)\,.
\end{equation}
where, if $\cV$ has infinite dimension, we assume the existence of the supertrace 
for the previous operator.
Indeed, the relation (\ref{univRLL-XX}) implies that the transfer matrices for
different values of the spectral parameter commute
\begin{equation} \label{commuting}
[t_{1\ldots L}(\lambda)\,,\,t_{1\ldots L}(\mu)]=0\,.
\end{equation}
Here the cyclicity of the supertrace has been used.

Since the R-matrix is regular (namely it is a permutation in $\lambda=0$), logarithmic
derivatives in $\lambda=0$ give local operators. We choose the first one as 
XX-Hamiltonian
\begin{eqnarray}
H&=&t_{1\ldots L}(0)^{-1}\, \frac{dt_{1\ldots L}}{d\lambda}(0) \label{eq:univXXHam}\\
&=&\sum_{j=1}^{L} H_{j,j+1}\mb{with} H_{j,j+1}=P_{j,j+1}\,\Sigma_{j,j+1}
\nonumber
\end{eqnarray}
where we have used periodic boundary conditions, i.e. identified the site
$L+1$ with the first site. 
After (\ref{commuting}), we see that any expansion of the transfer matrix in the spectral
parameters $\lambda,\:\mu$ generates a set of commuting operators. In particular they commute
with the Hamiltonian (\ref{eq:univXXHam}), so are conserved charges. This formally
proves that the system is integrable.

Explicitly, the two sites Hamiltonian corresponding to the example (\ref{example}) reads
\begin{equation}
H_{j,j+1} = \sum_{i\neq N, N+M~} \sum_{j=N,N+M} 
\Big( (-1)^{[j]}\,E_{ij} \otimes E_{ji} + (-1)^{[i]}\,E_{ji} \otimes E_{ij}
\Big)\,.
\end{equation}

\subsection{Symmetries of the universal XX models}
The choice of the fundamental projectors in (\ref{def:univpi}) directly 
fixes the symmetries of the model.

One easily shows that an operator $\MM\in \mbox{End}(\cW)\oplus \mbox{End}(\wt{\cW})$ 
commutes with the projectors (\ref{def:univpi}); then it commutes with the R-matrix in the
following sense
\begin{equation}
(\MM_{1}+\MM_{2})\,R_{12}(\lambda) = 
R_{12}(\lambda)\,(\MM_{1}+\MM_{2})\,.
\label{eq:symunivRxx}
\end{equation}
Commutation does not hold if the operator mixes the two subspaces.

As a consequence of (\ref{eq:symunivRxx}), the transfer matrix also has a
symmetry (super)algebra 
\be
\cS=\mbox{End}(\cW)\oplus \mbox{End}(\wt{\cW})
\ee 
with generators given by 
\begin{equation}\label{generasimm}
\MM_{<1\ldots L>}=\MM_{1}+\MM_{2}+\ldots+\MM_{L}\,.
\end{equation}
The same is true for any Hamiltonian $H$ built from the transfer matrix so (\ref{generasimm})
commute with the Hamiltonian\footnote{In principle, this construction cannot exclude
the existence of operators that commute with the Hamiltonian but not with the R-matrix. 
In that case, these additional symmetries would have the strange feature of 
not being symmetries of at least one conserved charge (by reconstructing the R-matrix 
from an expansion).}.

We can reverse this construction: we require a symmetry algebra $\cS$ from which 
we construct the subspaces $\cW$ and $\wt{\cW}$. This uniquely fixes 
the fundamental projector $\pi$ that immediately leads to obtain the XX model
possessing $\cS$ as symmetry.

The example (\ref{example}) admits $\cS=gl(N-1|M-1)\oplus gl(1|1)$ as symmetry
superalgebra whose generators $\MM$ have the form
\begin{equation}
\begin{array}{l@{~~~\mbox{for}~~~}c}
\displaystyle E_{jk}\ ,\ j,k\neq N,N+M  &  gl(N-1|M-1)\\
\displaystyle E_{jk}\ ,\ j,k=N,N+M &  gl(1|1).
\end{array}\label{eq:gen-glN-1}
\end{equation}


\section{Universal Hubbard models\label{sect:Hub}}

Starting with universal XX models, one can build universal 
Hubbard models, in the same way it has been done for usual and super 
Hubbard models \cite{EFGKK,DFFR1}. 
The logic will be to start from two possibly different universal XX models of section~\ref{sect:XX} and "glue" them with the generalisation of the construction given in section~\ref{sect:Hubbard}.

\subsection{R-matrix} 
We start with the $R$-matrices of two universal XX models,
$R^{\uparrow}_{12}(\lambda)$ and $R^{\downarrow}_{12}(\lambda)$, 
 living in two different sets of  spaces that we label by
$\uparrow$ and $\downarrow$. Let us stress that the two XX models can be based on two
different (graded) vector spaces $\cV^{\uparrow}$ and 
$\cV^{\downarrow}$, with two different projectors
$\pi^{\uparrow}$ and $\pi^{\downarrow}$. 

The Hubbard model is constructed from the coupling of these two XX 
models. Its $R$-matrix has two spectral parameters $\lambda_1\,,\lambda_2$ and reads:
\begin{equation}
R_{12}(\lambda_{1},\lambda_{2}) =
R^{\uparrow}_{12}(\lambda_{12})\,R^{\downarrow}_{12}(\lambda_{12}) +
\frac{\sin(\lambda_{12})}{\sin(\lambda'_{12})} \,\tanh(h'_{12})\,
R^{\uparrow}_{12}(\lambda'_{12})\,C^{\uparrow}_{1}\,
R^{\downarrow}_{12}(\lambda'_{12})\,C^{\downarrow}_{1}
\label{R-XXfus}
\end{equation}
where $\lambda_{12}=\lambda_{1}-\lambda_{2}$ and $\lambda'_{12}=\lambda_{1}+\lambda_{2}$. 
Moreover, $h'_{12}=h(\lambda_{1})+h(\lambda_{2})$ and the choice of the function $h(\lambda)$
is fixed by the proof of the Yang-Baxter equation.
Indeed, when the function $h(\lambda)$ is given by $\sinh(2h)=U\, \sin(2\lambda)$
for some free parameter $U$, the R-matrix (\ref{R-XXfus}) obeys YBE:
\begin{eqnarray}
R_{12}(\lambda_{1},\lambda_{2})\, 
R_{13}(\lambda_{1},\lambda_{3})\, 
R_{23}(\lambda_{2},\lambda_{3}) 
&=& 
R_{23}(\lambda_{2},\lambda_{3})\, 
R_{13}(\lambda_{1},\lambda_{3})\, 
R_{12}(\lambda_{1},\lambda_{2})\,.
\end{eqnarray}
As remarked in \cite{DFFR1} the proof relies only on some intermediate properties
that are not affected by the choice of the fundamental projectors (\ref{def:univpi}).
The proof follows the steps of the original proof by Shiroishi
\cite{shiro}, in the same way it has been done for algebras in 
\cite{EFGKK}. 
Moreover, it was already noticed in \cite{EFGKK} that one can couple
two XX models based on different $gl(M)$ algebras: this naturally 
extends to general (graded) vector spaces $\cV$.

The given R-matrix is regular but non symmetric. It satisfies unitarity (we correct here
an inconsequential typo that occurred in eq. 3.4 of \cite{DFFR1}) in the form
\begin{equation}
R_{12}(\lambda_{1},\lambda_{2})\,R_{21}(\lambda_{2},\lambda_{1}) = 
\left( \cos^{4}(\lambda_{12})
-\Big(\frac{\sin(\lambda_{12})}{\sin(\lambda'_{12})}\,\tanh(h'_{12})\cos^2(\lambda'_{12})\Big)^{2}
\right)\,\II_{1}\otimes\II_{2}
\ee
where $\II_{i}=\II\su\otimes\II\giu\,.$

\subsection{Monodromy and transfer matrix} 
We use the construction given in section~\ref{transferXX} to obtain 
the Hamiltonian of the system, starting with the `reduced' monodromy matrix
\begin{equation}\label{redmonodromy}
\cL_{0<1\ldots L>}(\lambda)=R_{01}(\lambda,\mu)\ldots R_{0L}(\lambda,\mu) \Big|_{\mu=0}\,.
\end{equation}
Any other choice for $\mu$ is possible but, at least in view of obtaining a local Hamiltonian, 
they do not give new information.
Provided the supertrace exists, the transfer matrix is given by
$$
t_{1\ldots L>}(\lambda)=\str_{0}\cL_{0<1\ldots L>}(\lambda)
$$
Then, one gets
\begin{eqnarray}
 [H, t(\lambda)] = 0\ ,\quad \forall \lambda\ , \mb{for}
H = H(0)=t(0)^{-1}\,t'(0) 
\end{eqnarray}
The `reduced' R-matrices that enter in (\ref{redmonodromy}) take a particularly simple 
factorised form
\begin{equation}
R_{12}(\lambda,0) = 
\,R^{\uparrow}_{12}(\lambda)\,R^{\downarrow}_{12}(\lambda)\,I^{\uparrow\downarrow}_{1}(h)
\end{equation}
where
\begin{equation}
I^{\uparrow\downarrow}_{1}(h) = \II\otimes\II
+\tanh(\frac{h}{2})\,C^{\uparrow}_{1}\,C^{\downarrow}_{1}
\end{equation}
and we arrive at a Hubbard-like Hamiltonian
\begin{equation} \label{eq:HubHam}
H = \sum_{j=1}^{L}H_{j,j+1} = \sum_{j=1}^{L} \Big[
\Sigma^{\uparrow}_{j,j+1}\,P^{\uparrow}_{j,j+1}
+\Sigma^{\downarrow}_{j,j+1}\,P^{\downarrow}_{j,j+1}
+U\,C^{\uparrow}_{j}\,C^{\downarrow}_{j} \Big]
\end{equation}
where we have used periodic boundary conditions.

\subsection{Symmetries} 
The transfer matrix of generalized Hubbard  models admits
as symmetry (super)algebra the direct sum of the symmetry algebras of the 
XX components
\be\label{simmetria}
\cS=\mbox{End}(\cW^{\uparrow})\oplus \mbox{End}(\wt{\cW}^{\uparrow})
\oplus \mbox{End}(\cW^{\downarrow})\oplus \mbox{End}(\wt{\cW}^{\downarrow})\,.
\ee
To prove this symmetry, it is useful to remark that (\ref{eq:symunivRxx}) 
can be now specialised to the cases up and down. Moreover, the up R-matrix commutes with
the down generators and viceversa. 
We also check that
\begin{equation}
\MM\,C^{\sigma}=C^{\sigma}\,\MM\,, \qquad \sigma=\uparrow, \downarrow
\end{equation}
where 
\be
\MM=\MM^{\uparrow}+\MM^{\downarrow}  \qquad \mbox{and} \qquad 
\MM^\sigma\in \mbox{End}(\cW^{\sigma})\oplus \mbox{End}(\wt{\cW}^{\sigma})\,.
\ee
Thus, one gets 
\begin{equation}
[R_{12}(\lambda,0)\,,\, \MM^{\uparrow}_{1}+\MM^{\uparrow}_{2}]=0
=[R_{12}(\lambda,0)\,,\,\MM^{\downarrow}_{1}+\MM^{\downarrow}_{2}]
\end{equation}
that can be easily extended to hold for the monodromy and transfer matrices and for the
Hamiltonian; the generators of the symmetry have the form
\begin{equation}
\MM^{\uparrow}=\sum_{j=1}^{L}\MM^{\uparrow}_{j} \mb{and} 
\MM^{\downarrow} =\sum_{j=1}^{L}\MM^{\downarrow}_{j}
\end{equation}
The ordinary Hubbard case and all the cases where $\cV^{\sigma}$ is two dimensional
are special because, in addition to the list of generators contained in (\ref{simmetria}), 
there are new generators given by
\be
V^{\pm}=\sigma^{\pm}_{\uparrow}\otimes\sigma^{\pm}_{\downarrow}\,, \qquad
W^{\pm}=\sigma^{\pm}_{\uparrow}\otimes\sigma^{\pm}_{\downarrow}\,.
\ee
To be precise, $V^{\pm}$ commutes with the Hamiltonian if $L$ is even while 
$W^{\pm}$ commutes in all cases. 
These additional generators do not commute with $H$ if $\mbox{dim}(\cV^{\sigma})>2$; 
they are responsible for the $SU(2)\times SU(2)$ symmetry of the even Hubbard model.

\section{Perturbative expansion of the Hubbard-like Hamiltonian\label{sect:pert}}
We expand the Hamiltonian (\ref{eq:HubHam}) in the 
inverse coupling $\frac1{U}$; according to (\ref{couplings}), this corresponds to the 
small coupling expansion of the gauge theory. 
Indeed, precisely that expansion has been used 
in \cite{Rej:2005qt} to match the $SU(2)$ dilatation operator with the 
effective Hamiltonian of the Hubbard model. The system was taken at 
half-filling to guarantee the required spin chain behaviour.
With the form of the potential used in (\ref{oldHubb}) the half-filled condition is 
enforced by the $U\rightarrow\infty$ requirement itself. 

We take the set of all Hamiltonian eigenstates whose 
leading energy term is $-LU$, for large positive $U$. These states are selected 
by the following projector 
\begin{equation}
\Pi_{0}=\prod _j  \big(\pi^{\uparrow}_j-\pi_j^{\downarrow}\big)^2 = 
\prod _j  \big(\wt{\pi}^{\uparrow}_j-\wt{\pi}_j^{\downarrow}\big)^2=
\Pi_{0}^2\,.\label{eq:Pi0}
\end{equation}
that projects on the subspace where, on each site, one and only one among 
$\wt\pi^{\uparrow}_j \,,\ \wt\pi_j^{\downarrow}$ has nonzero action. 

We follow the method introduced by Klein and Seitz \cite{klstz} to obtain an effective 
Hamiltonian for the corrections to the leading energy $-LU$:
\begin{equation}\label{ks}
H_{\text{eff}}=\frac1{U} H_{\text{eff}}^{(2)}+\frac1{U^3} 
H_{\text{eff}}^{(4)}+ \ldots
\end{equation}
For $L>2$ the second order effective Hamiltonian is 
\begin{equation}\label{secondord}
H_{\text{eff}}^{(2)}=\sum _j H_{\text{eff}\ j,j+1}^{(2)} =
2\sum _j \big(1+P^{\uparrow}_{j,j+1} P^{\downarrow}_{j,j+1}\big) 
\big(\pi^{\uparrow}_j\,\wt\pi_j^{\downarrow}\,
\wt\pi^{\uparrow}_{j+1}\,\pi^{\downarrow}_{j+1}+
\wt\pi^{\uparrow}_j\, \pi_j^{\downarrow}\,
\pi^{\uparrow}_{j+1}\, \wt\pi^{\downarrow}_{j+1}\big) 
\end{equation}
For the ordinary Hubbard model this expression can be given in terms of Pauli matrices
\be 
H_{\text{eff}}^{(2)}=\sum_{i=1}^L(1-\mbox{\boldmath$\sigma$}_i\mbox{\boldmath$\sigma$}_{i+1})
\ee
where the fermionic oscillators of (\ref{oldHubb}) have disappeared and only spin 
degrees of freedom are left ($\mbox{\boldmath$\sigma$}=(\sigma^x,\sigma^y,\sigma^z)$). 

The structure of the two-sites Hamiltonian $H_{\text{eff}\ i,i+1}^{(2)}$ can be obtained 
explicitly. In matricial form, it has diagonal block structure, with blocks given by
one of the two matrices 
\begin{equation}\label{twosites5}
B_{-}=\begin{pmatrix} 1& -1\\-1 &1 \end{pmatrix} \qquad \text{or}
\qquad B_{+}=\begin{pmatrix} 1& 1\\1 &1 \end{pmatrix} \,,
\end{equation}
all other entries being zero.
The number of appearances of each block depends on the actual model under examination.

\section{Jordan-Wigner transformation \label{sect:JW}}

Let us consider $p$ sets of fermionic oscillators $c_{i}^{(q)},
c_{i}^{(q)\dagger}$ ($i=1,\ldots,L$ and $q=1,\ldots,p$) that satisfy the
usual anticommutation relations
\begin{equation}
\{ c_{i}^{(q)},c^{(q')\dagger}_{j} \} = \delta_{ij} \, \delta_{qq'} \qquad \{
c_{i}^{(q)},c_{j}^{(q')} \} = \{ c^{(q)\dagger}_{i},c^{(q')\dagger}_{j} \} =
0
\end{equation}
One defines the following matrix (where $n_{i}^{(q)} = c_{i}^{(q)\dagger}
c_{i}^{(q)}$ is the usual number operator)
\begin{equation}
X^{(q)}_i = \left( 
\begin{array}{cc}
1-n_{i}^{(q)} & c_{i}^{(q)} \\
c_{i}^{(q)\dagger} & n_{i}^{(q)} \\
\end{array}
\right)
\end{equation}
The entries $X_{i;\alpha\beta}^{(q)}$ of this matrix have a natural gradation
given by $[\alpha]+[\beta]$ where $[1]= 1$ and $[2] = 0$.

In the $gl(2^{p-1}|2^{p-1})$ case, one defines at each site $i$ the
generators
\begin{equation}
X_{i;\alpha_1\ldots\alpha_p,\alpha'_1\ldots\alpha'_p} = (-1)^{s} \,
X_{i;\alpha_1\alpha'_1}^{(1)} \; \ldots \; X_{i;\alpha_p\alpha'_p}^{(p)}
\;\; \mbox{where} \;\;
s = \sum_{a=2}^{p} [\alpha_a] \Big( \sum_{b=1}^{a-1} \big( [\alpha_b] +
[\alpha'_b] \big) \Big)
\end{equation}
It is easy to verify the following properties:
\begin{eqnarray}
&& \big( X_{i;\alpha_1\ldots\alpha_p,\alpha'_1\ldots\alpha'_p} \big)^\dagger =
X_{i;\alpha'_1\ldots\alpha'_p,\alpha_1\ldots\alpha_p} \\
&& X_{i;\alpha_1\ldots\alpha_p,\alpha'_1\ldots\alpha'_p} \;
X_{i;\beta_1\ldots\beta_p,\beta'_1\ldots\beta'_p} = 
\delta_{\alpha'_1\beta_1} \ldots \delta_{\alpha'_p\beta_p} \,
X_{i;\alpha_1\ldots\alpha_p,\beta'_1\ldots\beta'_p} \\
&& \sum_{\alpha_1,\ldots,\alpha_p}
X_{i;\alpha_1\ldots\alpha_p,\alpha_1\ldots\alpha_p} = 1 \\
&& X_{i;\alpha_1\ldots\alpha_p,\alpha'_1\ldots\alpha'_p} \;
X_{j;\beta_1\ldots\beta_p,\beta'_1\ldots\beta'_p} = (-1)^{g}
X_{j;\beta_1\ldots\beta_p,\beta'_1\ldots\beta'_p} \;
X_{i;\alpha_1\ldots\alpha_p,\alpha'_1\ldots\alpha'_p} \qquad (i \ne j) \qquad \\
&& \mbox{where} \;\; g = \Big(\sum_{a=1}^p \big([\alpha_a] + [\alpha'_a]\big)\Big)
\Big(\sum_{b=1}^p \big([\beta_b] + [\beta'_b]\big)\Big) \nonumber
\end{eqnarray}
This means that the operators $X_{i;\alpha_1\ldots\alpha_p,\alpha'_1\ldots\alpha'_p}$
built out of fermionic oscillators are actually a realisation of the $gl(2^{p-1}|2^{p-1})$
superalgebra. 
A generic case $gl(N|M)$ can be understood as contained in the smallest superalgebra
for which $N,M<2^{p-1}$. The unwanted states can be consistently projected out.

\section{Examples\label{sect:examples}} 
It is possible to construct examples of both XX and Hubbard-like Hamiltonians. Clearly, 
the XX ones are "quasi-free models" because they do not contain external potentials
and, if written with fermionic oscillators, they only contain hopping terms.
In spite of this, they show curious "screening effects" namely particles that are allowed
to move only if particles of other types are present (or absent, depending on the case).
We will concentrate on universal Hubbard model examples.

The first example to cite is, of course, the original Hubbard model of 
section~\ref{sect:Hubbard}, that is described in this formalism as 
$gl(1|1)\oplus gl(1|1)$ with the choice $\pi^{\uparrow}=\pi^{\downarrow}=E_{11}$,
$\wt{\pi}^{\uparrow}=\wt{\pi}^{\downarrow}=E_{22}$.

\subsection{$gl(2|2)\oplus gl(2|2)$ Hubbard Hamiltonian }
This is a more complete example of the models under examination. It precisely implements 
two copies (up and down) of the example (\ref{example}) with $N=M=2$. 
The kinetic term of the Hamiltonian has a factorised form 
\begin{eqnarray}\label{hamilt}
H_{\text{Hub}} &\!\!=\!\!& \sum_{i=1}^{L} \; \Big\{ \;
\sum_{\sigma=\uparrow,\downarrow} \big( \cd{\sigma,i} \cc{\sigma,i+1}
+ \cd{\sigma,i+1} \cc{\sigma,i} \big)
\big( c_{\sigma,i}'^\dagger c_{\sigma,i+1}' + c_{\sigma,i+1}'^\dagger c_{\sigma,i}' + 1 
- n_{\sigma,i}' - n_{\sigma,i+1}' \big) \nonumber \\
&& + \, U (1-2n_{\uparrow,i})(1-2n_{\downarrow,i}) \; \Big\}
\end{eqnarray}
where the factor 
\be
\mathcal{N}'_{\sigma,i,i+1}=\big( c_{\sigma,i}'^\dagger
c_{\sigma,i+1}' + c_{\sigma,i+1}'^\dagger c_{\sigma,i}' + 1 - n_{\sigma,i}'
- n_{\sigma,i+1}' \big)
\ee
multiplies an ordinary Hubbard hopping term ; only unprimed particles enter into the potential.
There are four types of fermionic particles, respectively generated by 
$\cd{\uparrow,i}\,,\cd{\downarrow,i}\,,c_{\uparrow,i}'^\dagger \,,c_{\downarrow,i}'^\dagger$
so that they define a 16 dimensional vector space on each site. 
The corresponding numbers of particles are conserved.

The factor $\mathcal{N}'_{\sigma,i,i+1}$ works on a $4\times 4$ space
and its eigenvalues are $\pm 1$ with two-fold multiplicity. In particular this means that 
it cannot vanish, $\mathcal{N}'_{\sigma,i,i+1}\neq 0$.
Moreover, if no primed particles are present, $\mathcal{N}'_{\sigma,i,i+1}=1\,, ~\forall ~\sigma,i$. 
The same is true if the system is fully filled with primed particles
in which case $\mathcal{N}'_{\sigma,i,i+1}=-1$ therefore two of the sectors 
described by this Hamiltonian are equivalent to the ordinary Hubbard model. A Russian doll 
structure is appearing: if the projectors are well chosen, a larger model contains
the small ones.

If there are primed particles only, the energy vanishes (but not momentum). 
If the potential is interpreted as a Coulomb repulsion, then unprimed particles only  
carry electric charge. 

The compound objects formed by 
~$\cd{\sigma,i}\,c_{\sigma,i}'{^\dagger}$~ are rigid: no other term in the Hamiltonian
can destroy 
them. In this sense, we have four types of carriers, with the same charge but different behaviours:
two are the elementary objects $\cd{\sigma,i}$ in two polarisations 
$\sigma=\uparrow\,,\downarrow$, 
two are the compound objects (in two polarisations).

The symmetry, according to (\ref{simmetria}), is 
$gl(1|1)\oplus gl(1|1)\oplus gl(1|1)\oplus gl(1|1)$. 

At second order in $\frac1{U}$ the following effective Hamiltonian appears
\be\label{secondo}
H_{\textrm{eff}}^{(2)} = - \frac{1}{U}\sum_{i=1}^{L}  \left[ (\frac12 -
2 S_{i}^z S_{i+1}^z) - (S_{i}^+ S_{i+1}^- + S_{i}^- S_{i+1}^+) \,
\mathcal{N}_{\uparrow,i, i+1}' \, \mathcal{N}_{\downarrow,i, i+1}' \right]
\ee
that looks like a deformation of an XXX model. 
It has an enhancement of symmetry with respect to (\ref{hamilt}) in the sense that 
its symmetry is $gl(2|2)\oplus gl(2|2)$.

The two sites action of (\ref{secondo}) is a $64\times 64$ matrix that can be easily 
disentangled leading to both the blocks given in (\ref{twosites5}). 
In summary, it has eigenvalues 0 and 2, ~$0$~ with multiplicity 48, ~$2$~ with multiplicity 16.

\subsection{$gl(4)\oplus gl(4)$ Hubbard Hamiltonian}
We consider the model based on $gl(4)\oplus gl(4)$ and take the projectors according to 
the example (\ref{example}) with $N=4\,, ~M=0$, in two copies (up and down)
\begin{eqnarray}
H_{\text{Hub}} &\!\!=\!\!& \sum_{i=1}^{L} \; \Big\{ \;
\sum_{\sigma=\uparrow,\downarrow} \big( 
\cd{\sigma,i} \cc{\sigma,i+1} c_{\sigma,i}'^\dagger c_{\sigma,i+1}'
+ \cd{\sigma,i+1} \cc{\sigma,i} c_{\sigma,i+1}'^\dagger c_{\sigma,i}'+
\nonumber \\
&& + \, n_{\sigma,i}' n_{\sigma,i+1}' (\cd{\sigma,i}
\cc{\sigma,i+1}+\cd{\sigma,i+1} \cc{\sigma,i}) + n_{\sigma,i}
n_{\sigma,i+1} (c_{\sigma,i}'^\dagger
c_{\sigma,i+1}'+c_{\sigma,i+1}'^\dagger c_{\sigma,i}') \big)+ \nonumber \\
&& + \, U
(1-2n_{\uparrow,i}n_{\uparrow,i}')(1-2n_{\downarrow,i}n_{\downarrow,i}') \;
\Big\}\,.\label{hamilt4}
\end{eqnarray}
This model has the same vector space dimension of the $gl(2|2)$ one (\ref{hamilt}), 
$\mbox{dim}(\cV)=16$ but the elementary projectors are different and lead to slightly different 
interactions. That the two Hamiltonians are different is manifest if one examines the 
original form (\ref{eq:HubHam}) with the basis matrices $E_{\alpha \beta}$, before the Jordan-Wigner transformation.

Here there is complete symmetry between primed and non-primed particles; 
the effect of Coulomb repulsion only appears when both primed and unprimed particles 
are on the same site; if one of these types is alone, no Coulomb interaction is felt.
Observe that if $n'_{\sigma,i}=1$ everywhere and for all polarisations (or else if
$n_{\sigma,i}=1$), we re-obtain the $gl(1|1)$ Hubbard model.
The kinetic term also has a strange feature: a particle is allowed to move only if it
is accompanied by a particle of the same polarisation (i.e. up with up) but opposite
type (i.e. primed with unprimed).

\subsection{$gl(4|4)\oplus gl(4|4)$ Hamiltonian}
Following the example (\ref{example}), the following Hubbard Hamiltonian is obtained
\begin{eqnarray}
H_{Hub}^{gl(4|4)} &\!\!=\!\!& \sum_{i=1}^{L} \; \Big\{ \;
\sum_{\sigma=\uparrow,\downarrow} \big( c_{\sigma,i}^\dagger c_{\sigma,i+1}
+ c_{\sigma,i+1}^\dagger c_{\sigma,i} + 1 - n_{\sigma,i} - n_{\sigma,i+1}
\big) \Big( c_{\sigma,i}'^\dagger c_{\sigma,i+1}' c_{\sigma,i}''^\dagger
c_{\sigma,i+1}'' \nonumber \\
&& + \; c_{\sigma,i+1}'^\dagger c_{\sigma,i}' c_{\sigma,i+1}''^\dagger
c_{\sigma,i}'' - n_{\sigma,i}' n_{\sigma,i+1}' (c_{\sigma,i}''^\dagger
c_{\sigma,i+1}'' + c_{\sigma,i+1}''^\dagger c_{\sigma,i}'') \nonumber \\
&& - \; n_{\sigma,i}'' n_{\sigma,i+1}'' (c_{\sigma,i}'^\dagger
c_{\sigma,i+1}' + c_{\sigma,i+1}'^\dagger c_{\sigma,i}') \Big) + U
(1-2n_{\uparrow,i}'n_{\uparrow,i}'')
(1-2n_{\downarrow,i}'n_{\downarrow,i}'') \; \Big\} 
\end{eqnarray}
Here there are six types of fermions, $\cd{\sigma}, c'^\dagger_{\sigma}, c''^\dagger_{\sigma}$
so the local (one site) space of states is $64\times 64$. 

One observes that this Hamiltonian exhibits a `Russian doll' structure.
Indeed, there are four sectors in the space of states where the $gl(4|4)$
Hamiltonian reduces to the $gl(2|2)$ one, that also reduces to the $gl(1|1)$ one.
For example, one sector is given by 
$n_{\uparrow,i}'' = n_{\downarrow,i}' = 1$ for $1 \le i \le L$.

\section{Two-particles interaction\label{twopart}}
We sketch here the preliminary effects that we observed studying two particles in interaction.
It is convenient to consider a reference state as being a particle "vacuum" (pseudovacuum)
\be
\Omega= \mathop{(e_1^{\uparrow}\otimes e_1^{\downarrow})}_{\scriptscriptstyle 1} \otimes
\mathop{(e_1^{\uparrow}\otimes e_1^{\downarrow})}_{\scriptscriptstyle 2} \otimes \dots 
\mathop{(e_1^{\uparrow}\otimes e_1^{\downarrow})}_{\scriptscriptstyle L}
\ee
where index under the tensor product symbol labels the lattice sites.
All other states are considered excitations of this pseudovacuum. 
Then particles are distinguished by the type, according
to the subspaces $\cW,~\wt{\cW}$:
\begin{eqnarray}
a,b,\dots &\in & \cW  \nonumber \\
\tilde{a},\tilde{b},\dots &\in & \wt{\cW} \nonumber
\end{eqnarray}
and an upper index $\uparrow, \downarrow$ will be added to distinguish polarisation\footnote{As already remarked, notice that particles of different polarisation or different type are not to
be understood as conjugated: for example, $a\su$ and $a\giu$ are different objects}.

Within the universal XX models, all particles satisfy the exclusion principle, namely they 
cannot appear on the same site. If two particles are both from $\cW$ or both from 
$\wt{\cW}$, they reflect each other; if they are one from $\cW$, one from $\wt{\cW}$,
they traverse each other by tunnel effect.

In the universal Hubbard models, the coupling activates a sort of electrostatic interaction 
felt by particles of opposite polarisation only. Indeed, the potential term in (\ref{eq:HubHam})
squares to the identity (\ref{eq:opC}) so on one site states it has eigenvalues $\pm U$. 
Which sign occurs is dictated by the membership to $\cW$ or $\wt{\cW}$ according to the rule:
with $U>0$, equal type particles $a\su a\giu$ or $\tilde{a}\su \tilde{a}\giu$ repel each other
but different type particles $a\su \tilde{a}\giu$ or $\tilde{a}\su a\giu$ attract each other.
Observe that the vacuum itself is in the repulsive case so actually the only 
"visible" effect is the attractive one.

\section{Conclusions}
We have constructed universal XX and Hubbard model Hamiltonians based on general properties of 
projectors and permutations. The underling algebraic structure could be an ordinary or graded algebra $gl(N|M)$ or possibly and infinite dimensional algebra.
We have full control of the symmetries of the models and we have performed the
perturbative calculation \emph{\`a la} Klein and Seitz \cite{klstz} in the large coupling limit.

We have emphasised that the gradation makes the Jordan--Wigner transformation a local
isomorphism. Therefore, the interpretation of the graded models in terms of `electrons' 
is more natural.

We discussed some examples, with their phenomenology. There, 
it would be very nice to see if the major screening effects observed (\ref{hamilt}) and 
(\ref{hamilt4}) in the Hamiltonians can be interpreted in some condensed matter context.

The next step in the study of our models is the determination of the spectrum
and of the Bethe equations, as they were constructed for Hubbard or
generalisation, using the algebraic Bethe ansatz \cite{YueDegu,RamMar,martins,Hubsu4}
and the coordinate Bethe Ansatz of Lieb-Wu \cite{LiebWu}. 
This is an heavy calculation which we
postpone for further publication, but from the analytical Bethe ansatz
approach, one can guess their form. In particular, as for spin chain
models, one expects as many presentations of the Bethe equations as there
are inequivalent Dynkin diagrams. All these presentations should lead to
the same spectrum. For more informations, we refer to \cite{selene,RS} where
similar calculations were performed in the case of XXX super spin chains.

Our models are graded by construction so they naturally contain bosonic as well fermionic 
degrees of freedom. We are working on examples with bosonic particles, that necessarily will be 
on infinite dimensional algebras.

Finally, the Bethe equations will allow us to keep in touch with super-symmetric gauge theories,
where integrability appears precisely in relation to the Hubbard model.

\section*{Acknowledgments}
GF thanks INFN for a post-doctoral fellowship and for financial support.
This work was partially supported by the EC Network
`EUCLID. Integrable models and applications: from strings to condensed
matter', contract number HPRN-CT-2002-00325 and by the ANR project,
`Theories de jauge superconformes', number BLAN06-3 143795.


\end{document}